\documentclass[journal]{IEEEtran}
\usepackage{amssymb,amsfonts,amsthm,bm}
\usepackage{array}
\usepackage{amsmath} % for bmatrix
\usepackage{upgreek} %\upalpha, \upbeta, ... \upeta, .. \Upalpha, \Upbeta ...
\usepackage{graphicx}
\usepackage{verbatim} % for \begin{comment} ~ \end{comment}
\newcommand{\abs}[1]{\left\vert#1\right\vert}

\newcommand{\lp}{\left(}
\newcommand{\rp}{\right)}
\newcommand{\noi}{\noindent}
\newcommand{\non}{\nonumber}
\newcommand{\teq}{\triangleq} % required {amssymb} package
\newcommand{\eps}{\epsilon}

\newcommand{\dz}{{\dot z}}

\newcommand{\vtab}{\vspace{2mm}}

\newcommand{\sgn}{\,{\rm sgn}}
\newcommand{\sat}{\,{\rm sat}}

\newcommand{\da}{{\dot a}}
\newcommand{\dal}{{\dot \alpha}}
\newcommand{\dsg}{{\dot \sigma}}
\newcommand{\eal}{e_\alpha}
\newcommand{\esg}{e_\sigma}
\newcommand{\deal}{{\dot{e}_\alpha}}
\newcommand{\desg}{{\dot{e}_\sigma}}

\newcommand{\Ld}{{L_\delta}}
\newcommand{\eda}{{e^d_\alpha}}
\newcommand{\tdi}{{t^i_\delta}}

\begin{document}
%\title{Uniformly Convergent Differentiator with Prescribed Accuracy}
\title{Switching Differentiator}
%\title{Estimation of the Time-Derivatives of a Signal using Switching Differentiator}
\author{Jang-Hyun~Park%,~%\IEEEmembership{Member,~IEEE,}
%        Seong-Hwan Kim,~%\IEEEmembership{Member,~IEEE,}
%        and~Tae-Sik Park%~\IEEEmembership{Member,~IEEE,}% <-this % stops a space
\thanks{J.-H. Park is with the Department of Electrical and Control Engineering, Mokpo National University, Chonnam 58554, Korea (e-mail: jhpark72@mokpo.ac.kr }% <-this % stops a space
}


\markboth{Journal of \LaTeX\ Class Files,~Vol.~14, No.~8, August~2015}%
{Shell \MakeLowercase{\textit{et al.}}: Bare Demo of IEEEtran.cls for IEEE Journals}



\maketitle

\begin{abstract}
A novel switching differentiator that has considerably simple form is proposed. Under the assumption that time-derivatives of the signal are norm-bounded, it is shown that estimation errors are convergent to the zeros asymptotically. The estimated derivatives shows neithor chattering nor peaking pheonomenon. A 1st-order diffentiator is firstly proposed and, by connecting this differentiator in series, higher-order derivatives are also available. Simulation results show that the proposed differentiator show extreme performance compared to the widly used previous differntiators such as high-gain observer or hige-order sliding mode differentiator.
\end{abstract}

% \fi
\IEEEpeerreviewmaketitle
%%%%%%%%%\%%%%%%%%%\%%%%%%%%%\%%%%%%%%%\%%%%%%%%%\%%%%%%%%%\%%%%%%%%%\%%%%%%%%%
%
%
%
%
%
\section{Introduction}
%
%
%
%
%
%%%%%%%%%\%%%%%%%%%\%%%%%%%%%\%%%%%%%%%\%%%%%%%%%\%%%%%%%%%\%%%%%%%%%\%%%%%%%%%
On-line differentiator for a given signal is widely utilzed in control system containing PID regulators \cite{JHan09}, states observers \cite{HKKhal17,HKKhal17b,AEBrys69,ALevant03,ALeva09}, fault diagnosis schemes\cite{DEfim11}, and active disturbance rejection \cite{ZGao06,JHan09}. The performance of the differentiator is crucial since it is directly connected to that of the controller. To mention just a few, there are linear differentiator \cite{SCPei89}, high-gain observer (HGO) \cite{HKKhal17,HKKhal17b}, high-order sliding mode (HOSM) differentiator \cite{ALevant03,ALeva09}, the super-twisting second-order sliding-mode algorithm \cite{JDavi05}, uniformly convergent differerntiator \cite{MTAngu13}, singular perturbation technique based differentiator \cite{XWang07}, augmented nonlinear differentiator (AND) \cite{XShao17}, etc.\par
Among the various differentiators, HGO and HOSM differentiator are widly adopted in the controller design for nonlinear systems. The HGO whose dynamics is linear in the estimation error has a shortcoming of peaking due to nonzero initial condition. This also leads to the non-robust against measurement disturbance. The HOSM differentiator has the property of finite-time exact convergence. However, since it contains discontinuous switching function in its dynamics, the chatterings in its estimations are invevitable and its dynamics are rather complex. In the presented differentiators in \cite{JDavi05, XWang07, MTAngu13, XShao17}, their nonlinear dynamics become complicated which leads to numerical problems for the practical use as well as simulation come out. \par
In this paper, a novel switching differentiator (SD) that has considerably simple form is proposed. Under the assumption that time-derivative of the signal is norm-bounded, it is proven that estimation error is asymptotically convergent to the zero. The observed derivative shows neithor chattering nor peaking phenomenon. A 1st-order derivative estimator is firstly proposed and, by connecting the proposed differentiators in series, it is shown that higher-order derivatives are also available. Simulation results depict that the proposed differentiator shows extreme performace compared to other well-known differentiators.

%%%%%%%%%\%%%%%%%%%\%%%%%%%%%\%%%%%%%%%\%%%%%%%%%\%%%%%%%%%\%%%%%%%%%\%%%%%%%%%
%
%
%
%
%
\section {Main Result}
%
%
%
%
%
%%%%%%%%%\%%%%%%%%%\%%%%%%%%%\%%%%%%%%%\%%%%%%%%%\%%%%%%%%%\%%%%%%%%%\%%%%%%%%%
%-------------------------------------------------------------------------------
%
%
\subsection{Switching Differentiator}
%
%
%-------------------------------------------------------------------------------
Consider the time-varing signal $a(t)$ whose time derivative is to be estimated. Assume that $\abs{{\ddot a}(t)}\le L^*$ holds $\forall t>0$. The proposed SD has the following form
\begin{eqnarray} \label{sd}
\dal &=& k e_\alpha + \sigma \cr
\dsg &=& L\,{\rm sgn}( e_\alpha ).
\end{eqnarray}
where $\eal=a-\alpha$, $k$ is a positive design constant and $L$ is determined such that $L>L^*$. The $\alpha$ and $\sigma$ are expected to estimate $a$ and $\da$ respectively. The second error is denoted as
\begin{eqnarray} 
\esg &=& \da-\sigma \label{e1}
\end{eqnarray}
and whether this $\esg$ will converge to zero as time goes by is a main concern. Our main result is in the following theorem.\par
\vtab
{\it Theorem 1:} The $\sigma(t)$ in (\ref{sd}) is asymptotically convergent to $\dot a(t)$ .\par
\begin{figure}[!t]
	\centering
	\includegraphics[width=3.5in]{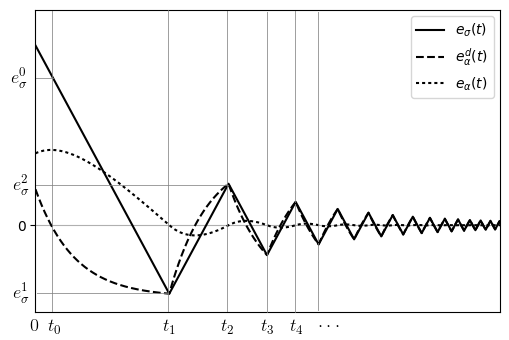}
	\caption{Typical trjectory of $\esg(t)$, $\esg^d(t)$ and $\eal(t)$.}
	\label{fig01}
\end{figure}

\vtab
{\it proof.} The time-derivatives of $\eal$ and $\esg$ are derived as
\begin{eqnarray} 
\deal &=& -k \eal + \esg \label{deal1}\\
\desg &=& {\ddot a}-  L\sgn(\eal). \label{desg1}
\end{eqnarray}
It is assumed the worst cast that the singal ${\ddot a}(t)$ acts to hinder the estimation as much as possible. This means that when $\eal(t)>0$, ${\ddot a}(t)$ maintains its extreme value $L^*$ which reduces the switching gain upmost. Conversly, if $\eal(t)$ is negative then ${\ddot a}(t)$ is assumed to maintain $-L^*$. Considering this assumption, (\ref{desg1}) can be redescribed as
\begin{eqnarray} \label{desg2}
\desg &=& -\Ld \, \sgn(\eal) \teq r(t)
\end{eqnarray}
where $\Ld = L-L^* (>0)$. From (\ref{deal1}) and (\ref{desg2}), the following dynamics is induced
\begin{eqnarray} \label{ddeal}
{\ddot e}_\alpha &=& -k \deal + \desg \cr
&=& -k \deal + r(t). 
\end{eqnarray}
Defining $\eda \teq \deal$, (\ref{ddeal}) becomes
\begin{equation} \label{deda}
\dot \eda = -k \eda + r(t)
\end{equation}
whose solution is
\begin{equation} \label{sol_eda}
\eda(t) = \eda(0) e^{-kt} + \int_0^t r(\tau) e^{-k(t-\tau)} d\tau. 
\end{equation}
In the case that $\eal(t)>0$ holds  for $0 \le t<t_1$, the solution for $\eda(t)$ is
\begin{eqnarray}
\eda(t) &=& \eda(0) e^{-kt} - \Ld \int_0^t e^{-k(t-\tau)} d\tau \cr 
&=& \eda(0) e^{-kt} - \frac{\Ld}{k} (1-e^{-kt}) \cr 
&=& \left( \eda(0) + \rho \right) e^{-kt} - \rho
\end{eqnarray}
since $r(t)=-\Ld$ for $t<t_1$ where $\rho=\frac{\Ld}{k}$. Because the first term decays exponetially, it is evident that there is $t_0(<t_1)$ such that $\eda(t)$ become negative for $t>t_0$. In the other case that the initial error $\eal(0)$ is negative, similar explanation is possible since $r(t)=\Ld$ and 
\begin{equation}
\eda(t) = \left( \eda(0) - \rho \right) e^{-kt} + \rho
\end{equation}
Note that the $t_0$ can be made arbitrarily small by increasing $\rho$. In either cases, there exist $t_0$ and $t_1$ ($0<t_0<t_1$) such that $\eal(t)$ turns its direction toward zero at $t=t_0$ and becomes zero at $t=t_1$. The typical trajectores of $\esg(t)$, $\eda (t)$ and $\eal (t)$ are illustrated in fig. \ref{fig01} for the convinience of the proof follows.\par
\begin{figure}[!t]
	\centering
	\includegraphics[width=3.7in]{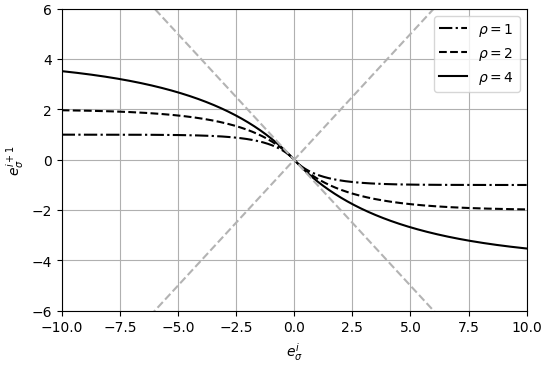}
	\caption{The graphs of $\esg^{i+1}$ versus $\esg^i$ with different $\rho$'s and $\esg^{i+1}= \pm \esg^{i}$ lines.}
	\label{fig02}
\end{figure}

\begin{figure}[!t]
	\centering
	\includegraphics[width=3.3in]{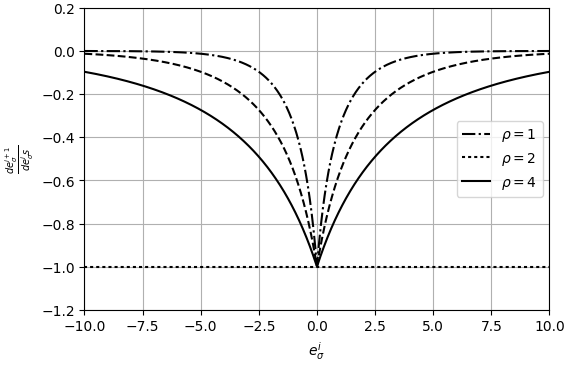}
	\caption{The graphs of $\frac{d\esg^{i+1}}{d\esg^i}$ with different $\rho$'s.}
	\label{fig03}
\end{figure}

\begin{figure}[!t]
	\centering
	\includegraphics[width=3.3in]{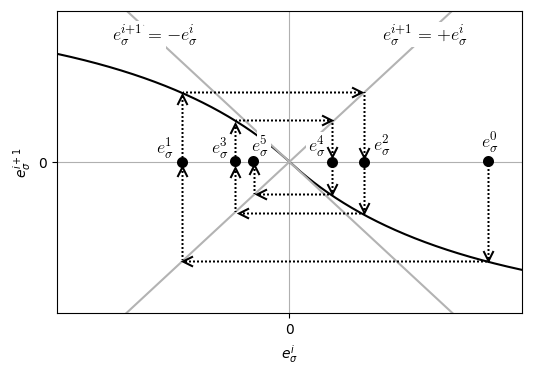}
	\caption{Typical variations of $\esg^i$'s by (\ref{esgip1}) for $i=0,1,2,3,4,5$. (black dots in $\esg^i$ axis)}
	\label{fig04}
\end{figure}

Let the time points that $\eal(t)=0$ holds be denoted as $t_i$, $(i=1,2,\cdots)$. It is evident from (\ref{deal1}) that $\eal(t_i)=0$ if and only if $\eda(t_i) = \esg(t_i)$. Since, in the time interval $t_i<t\le t_{i+1}$, $r(t)$ is constant (either $-\Ld$ or $\Ld$), the time solustions of $\eda(t)$ and $\esg(t)$ for $t_i<t\le t_{i+1}$ are
\begin{eqnarray}
\eda(t) &=& \esg(t_i)e^{-k(t-t_i)} + \frac{r}{k} \lp 1-e^{-k(t-t_i)} \rp \label{edat}\\
\esg(t) &=& \esg(t_i) + r(t-t_i). \label{esgt}
\end{eqnarray}
Here, $\eda(t_i) = k\eal(t_i)+\esg(t_i) = \esg(t_i)$ holds due to $\eal(t_i)=0$. The two right-hand sides of (\ref{edat}) and (\ref{esgt}) must be identical at $t=t_{i+1}$. Denoting $\tdi \teq t_{i+1} - t_i$ and $\esg^i = \esg(t_i)$ yields 
\begin{equation}
\esg^i e^{-k\tdi} + \frac{r}{k} \lp 1-e^{-k\tdi} \rp = \esg^i + r\tdi
\end{equation}
or
\begin{equation}\label{eq1}
\lp \esg^i - \frac{r}{k} \rp + r\tdi = \lp \esg^i - \frac{r}{k} \rp e^{-k\tdi}
\end{equation}
Note that the solution of $A+Bt=De^{-Ct}$ is $t=\frac{1}{C} W[\frac{CD}{B}e^{\frac{AC}{B}} ]-\frac{A}{B}$ where $W(\cdot)$ is Lambert-W function. Using this formula with 
\begin{eqnarray}
A &=& D = \lp \esg^i - \frac{r}{k} \rp \cr
B &=& r \cr
C &=& k
\end{eqnarray}
and defining
\begin{eqnarray}
x &\teq& \frac{CD}{B} = \frac{AC}{B}  \cr
&=& \frac{k}{r}\esg^i - 1  = -\frac{k}{\Ld}\abs{\esg^i} - 1 \cr
&=& -\frac{\abs{\esg^i}}{\rho} - 1 <-1
\end{eqnarray}
the solution of (\ref{eq1}) is
\begin{eqnarray}
\tdi &=& -\frac{\esg^i}{r} + \frac{1}{k}\lp 1+ W(x e^x) \rp 
\end{eqnarray}
Using this time interval, $\esg^{i+1} := \esg(t_{i+1})$ which is the function of $\esg^i$ can be obtained from (\ref{desg2}) as
\begin{eqnarray}\label{esgip1}
\esg^{i+1} &=& \esg^i + \int_{t_i}^{t_{i+1}}r dt \nonumber \\[2pt]
 &=& \esg^i + r \tdi \nonumber \\[3pt]
&=&  \frac{r}{k} \lp 1+ W(xe^x) \rp 
\end{eqnarray}
or
\begin{eqnarray}\label{esgip2}
\esg^{i+1} &=& -\sgn(\esg^i)\rho \lp 1+ W(xe^x) \rp.
\end{eqnarray}
The value of $W(xe^x)$ is $-1$ at $x=-1$ and approaches to $0^-$ as $x$ goes to $-\infty$. The graphs of $\esg^{i+1}$ versus $\esg^i$ with different $\rho$'s are illustrated in fig. \ref{fig02}.
%
%
%From (\ref{esgip1}) and fig. \ref{fig02}, it is evdient that $\abs{\esg^{i+1}} < \abs{\esg^{i}}$.
The slope of $\esg^{i+1}$ curve at $\esg^i=0$, which is denoted as $z$ in what follows, can be calculated from (\ref{esgip1}) as
\begin{eqnarray}
z  &\teq& \frac{d \esg^{i+1}}{d \esg^i} \Big\vert_{\esg^i=0} 
=  \frac{r}{k} \frac{d W(xe^x)}{d \esg^i} \Big\vert_{\esg^i=0} \label{z} \\
&=&  \frac{r}{k} \frac{e^xW(xe^x)}{xe^x} \Big\vert_{\esg^i=0} \frac{\esg^i}{1+W(xe^x)} \Big\vert_{\esg^i=0} \cr
&=&  \frac{r}{k} \frac{\esg^i}{1+W(xe^x)} \Big\vert_{\esg^i=0} \nonumber
\end{eqnarray}
since $x(\esg^i=0)=-1$ and $W(-e^{-1})=-1$. The numerator and denominator of the last term are all zeros. Thus, applying L'Hopital's theorem and derivating w.r.t $\esg^i$ both of them yields
\begin{eqnarray}
z &=&  \frac{r}{k} \frac{1}{ \frac{d \esg^{i+1}}{d \esg^i} \Big\vert_{\esg^i=0}}
=   \frac{r}{k} \frac{1}{ \frac{k}{r} z} \cr
&=&  \frac{1}{z}
\end{eqnarray}
where we used (\ref{z}). It is evident from $z<0$ (refer to fig. \ref{fig02}) that the valid solution of $z$ is $-1$. The graphs are illustrated in fig. \ref{fig03} for some $\rho$'s.
This means that the $\esg^{i+1}$ curve is exactly tangent to the $\esg^{i+1}=-\esg^i$ line at the origin and there are no crossing points between them except the origin. Thus, this results in asymptotical convergence of $\abs{\esg^{i}}$ to zero because $\abs{\esg^{i+1}} < \abs{\esg^{i}}$ for all $i>0$. The typical variations of $\esg^i$'s are illustrated in fig.\ref{fig04} as a black dot in $\esg^i$ axis.
This completes the proof.
\vtab \par
Note that the convergence properties hold globally and uniformly, which means that the errors converge to the zeros regardless of large initial gaps. The initial values of $\alpha$ and $\sigma$ can usually be chosen as zeros since the informations on $a(0)$ and $\da(0)$ may be difficult to obtain {\it a priori} in practice. It is worth noting that the proposed SD shows no peaking pheonomenon caused by nonzero initial errors during transient period. This property is very crucial because it leads to the robustness againt measurement noise.\par
\par
It is also worth noting that the effect of the discontinuous switching function $\sgn(\eal)$ is integrated and, therefore, the chattering in $\sigma$ is suppressed. In practice, it is hard to obtain the bounds on the norm-bound of ${\ddot a}(t)$, which leads to choose $L$ as a sufficiently large value. This is admittable since the chattering is suppressed by this reason, and it will be shown in simulations later.\par
%
%-------------------------------------------------------------------------------
%
%
\subsection{Higher-Order Differentiator}
%
%
%-------------------------------------------------------------------------------
The higher-order time derivatives of $a(t)$ can be available via series connection of (\ref{sd}). The series equations for $i=1,2,\cdots$ are
\begin{eqnarray} \label{ssd}
\dal_i &=& k e_{\alpha i} + \sigma_i \cr
\dsg_i &=& L\sgn(e_{\alpha i}) 
\end{eqnarray}
where $e_{\alpha i} = \sigma_{i-1}-\alpha_i$ with $\sigma_{0}=a(t)$. The $L$ is determined sufficiently large such that $L>\max_i L_i^*$ where $\abs{a^{{i+1}}(t)}\le L_i^*$. Then, the estimate of $a^{(i)}(t)$ is $\sigma_i(t)$ which is also expected to be asymptotically convergent.\par
%
%\vtab
%%
%{\it Theorem 2:} The $\sigma_i(t)$ in (\ref{sdn}) is asymptotically convergent to $a^{(i)}(t)$ .\par
%%
%%
%\vtab
%%
%The proof is trivial and omitted.
%%%%%%%%%\%%%%%%%%%\%%%%%%%%%\%%%%%%%%%\%%%%%%%%%\%%%%%%%%%\%%%%%%%%%\%%%%%%%%%
%
%
%
%
%
\section {Simulations}
%
%
%
%
%
%%%%%%%%%\%%%%%%%%%\%%%%%%%%%\%%%%%%%%%\%%%%%%%%%\%%%%%%%%%\%%%%%%%%%\%%%%%%%%%
In this section, using the proposed SD and other well-known differntiators, the time derivatives $\da(t)$ up to $a^{(4)}(t)$ are estimated through simulations where  of $a(t)=2\sin{t}+3\cos{3t}$. To compare the performaces with each other, the settling time of the esitmate of $a^{(4)}$ is deliberately set to about $0.1$s via tuning their design parameters. 
\subsection{Proposed SD}
The proposed series SDs (\ref{ssd}) is used to observe $\da$, $\ddot a$, $\dddot a$ and $a^{(4)}$. The whole formulas whose dynamic order is 8 is as follows.
\begin{eqnarray} \label{ssdn}
%\cr
\dal_1 &=& k e_{\alpha 1} + \sigma_1 {\rm~~~with~~} e_{\alpha 1} = a(t)-\alpha_1\cr
\dsg_1 &=& L\sgn(e_{\alpha 1}) 
%{\rm~with~}e_{\alpha 1} = a(t)-\alpha_1
\non \\[2pt]
%
% \cr
\dal_2 &=& k e_{\alpha 2} + \sigma_2 {\rm~~~with~~}e_{\alpha 2} = \sigma_1-\alpha_2 \cr
\dsg_2 &=& L\sgn(e_{\alpha 2})
%{\rm~with~}e_{\alpha 2} = \sigma_1-\alpha_2
\non \\[2pt] %\cr
%
% \cr
\dal_3 &=& k e_{\alpha 3} + \sigma_3 {\rm~~~with~~}e_{\alpha 3} = \sigma_2-\alpha_3 \cr
\dsg_3 &=& L\sgn(e_{\alpha 3}) 
\non \\[2pt] %\cr
%%
% \cr
\dal_4 &=& k e_{\alpha 4} + \sigma_4 {\rm~~~with~~}e_{\alpha 4} = \sigma_3-\alpha_4\cr
\dsg_4 &=& L\sgn(e_{\alpha 4})
\end{eqnarray}
%

\begin{figure}[!t]
	\centering
	\includegraphics[width=3.5in]{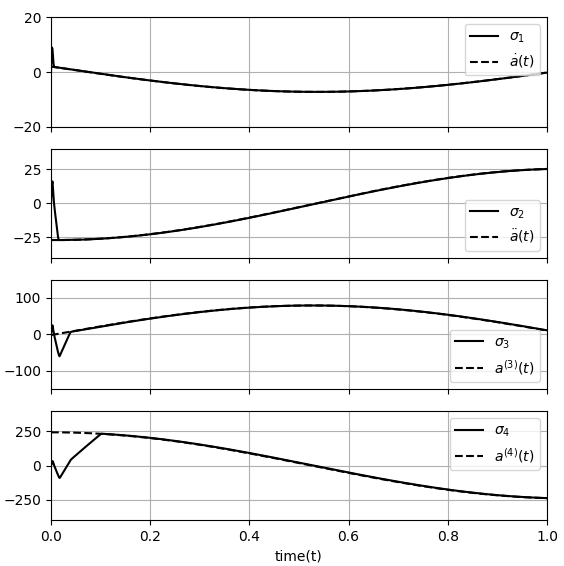}
	\caption{estimates of $\da(t)$, ${\ddot a}(t)$, ${\dddot a}(t)$, and $a^{(4)}(t)$ using proposed SSDs.}
	\label{sim11}
\end{figure}

%
\begin{comment}
%
\begin{eqnarray} \label{ssdn}
%e_{\alpha 1} &=& a(t)-\alpha_1\cr
\dal_1 &=& k (a-\alpha_1) + \sigma_1 \cr
\dsg_1 &=& k (a-\alpha_1) + L\tanh \left( (a-\alpha_1)/\eps \right)
%{\rm~with~}e_{\alpha 1} = a(t)-\alpha_1
\non \\[2pt]
%
%e_{\alpha 2} &=& \sigma_1-\alpha_2 \cr
\dal_2 &=& k (\sigma_1-\alpha_2) + \sigma_2 \cr
\dsg_2 &=& k (\sigma_1-\alpha_2) + L\tanh \left( (\sigma_1-\alpha_2)/\eps \right)
%{\rm~with~}e_{\alpha 2} = \sigma_1-\alpha_2
\non \\[2pt] %\cr
%
%e_{\alpha 3} &=& \sigma_2-\alpha_3 \cr
\dal_3 &=& k (\sigma_2-\alpha_3) + \sigma_3 \cr
\dsg_3 &=& k (\sigma_2-\alpha_3) + L\tanh \left(  (\sigma_2-\alpha_3)/\eps \right) 
%
\non \\[2pt] %\cr
%%
%e_{\alpha 4} &=& \sigma_3-\alpha_4 \cr
\dal_4 &=& k (\sigma_3-\alpha_4) + \sigma_4 \cr
\dsg_4 &=& k (\sigma_3-\alpha_4) + L\tanh \left( (\sigma_3-\alpha_4)/\eps \right)
\end{eqnarray}
%
%
\end{comment}
%
\noi The estimation of $a^{(i)}$ is $\sigma_i$ for $i=1,2,3,4$. The parameters are chosen as $k=3000$, $L=3000$. For the simulations, we used $\sat(\frac{e_{\alpha i}}{\eps})$ with $\eps=10^{-4}$ instead of $\sgn(e_{\alpha i})$ for $i=1,2,3,4$. Initial values of all the states in SDs are set to zeros. The result is in Fig. \ref{sim11}. Even in $\sigma_4(t)$ which is the estimation of $a^{(4)}(t)$ follows exactly to the real value after transient time of $0.1$ sec without any peaking or chattering.\par
\begin{figure}[!t]
	\centering
	\includegraphics[width=3.5in]{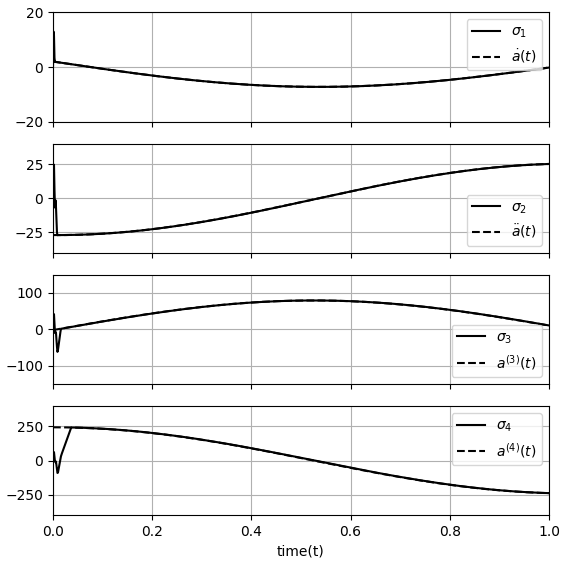}
	\caption{estimates of $\da(t)$, ${\ddot a}(t)$, ${\dddot a}(t)$, and $a^{(4)}(t)$ using proposed SSDs.}
	\label{sim12}
\end{figure}

Note that the transient time can be shortened further through increasing $k$ and $L$. In fig. \ref{sim12}, the simulation result is shown with parameters $k=5000$, $L=10000$. The settling time is shortened compared to fig. \ref{sim11}. 
It is worth to note that, in determining $L$, the upper bound of $a^{(5)}$ is not considered and it is chosen as sufficiently large value. For fairly large values of $k$ and $L$, the propoesed SDs show extreme performance while generating no peaking nor chattering in the estimated values at least in the simulations.
%
%-------------------------------------------------------------------------------
%
%
\subsection{HGO}
\begin{figure}[!t]
	\centering
	\includegraphics[width=3.5in]{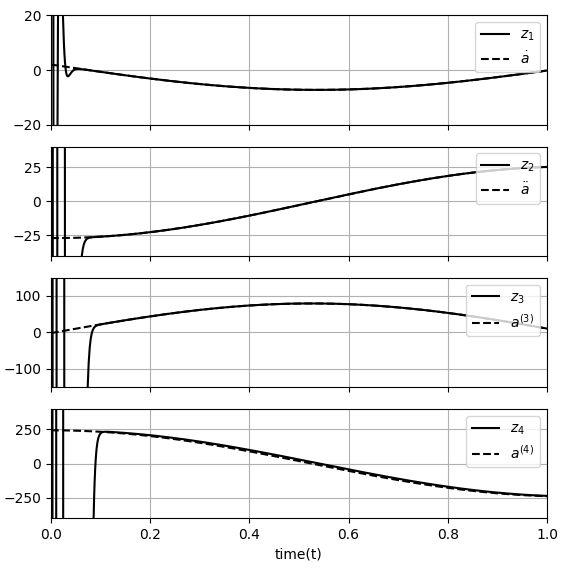}
	\caption{estimates of $\da(t)$, ${\ddot a}(t)$, ${\dddot a}(t)$, and $a^{(4)}(t)$ using HGO.}
	\label{sim21}
\end{figure}

%
%
%-------------------------------------------------------------------------------
The derivatives of the same signal using HGO \cite{HKKhal17} that has the following dynamics are esitmated.
\begin{eqnarray} \label{hgo}
%e &=& a-z_0 \cr
\dz_0 &=& z_1 + (c_0/\eps) (a-z_0) \cr
\dz_1 &=& z_2 + (c_1/\eps^2)(a-z_0) \cr
\dz_2 &=& z_3 + (c_2/\eps^3)(a-z_0) \cr
\dz_3 &=& z_4 + (c_3/\eps^4)(a-z_0) \cr
\dz_4 &=& (c_4/\eps^5)(a-z_0)
\end{eqnarray}
where $z_i$ ($i=1,2,3,4$) is the estimates of $a^{(i)}$. The parameters are determined such that the settling time of $z_4$ is about $0.1$s  as before. The determined parameters are  $c_0=47.5$, $c_1=902.5$, $c_2=8573.75$, $c_3 =40725.3125$, $c_4=77378.09375$ and $\eps=0.03$. The result is illustrated in Fig. \ref{sim21}. In this figure, the y-axis limits are identical to those of fig. \ref{sim11} for the convenience of comparison. Note that the peaking in transient period is enormous and grow rapidly as $i$ increases. In this simulation, the peak value for $z_1$ is $1552.1$ and about $4\times 10^9$ for $z_4$.
\subsection{HOSM differentiator}
\begin{figure}[!t]
	\centering
	\includegraphics[width=3.5in]{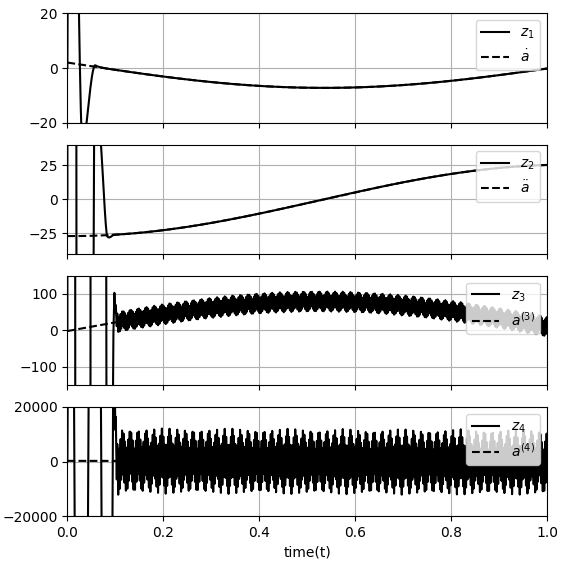}
	\caption{estimates of $\da(t)$, ${\ddot a}(t)$, ${\dddot a}(t)$, and $a^{(4)}(t)$ using HOSM differentiator.}
	\label{sim31}
\end{figure}

The HOSM differentiator \cite{ALevant03} for estimating up to $a^{(4)}$ has the following form.
\begin{eqnarray} \label{hgo}
%e &=& a-z_0 \cr
\dz_0 &=& -8L^{\frac{1}{5}} \lceil z_0-a \rfloor^{\frac{4}{5}} \teq v_0 \cr
\dz_1 &=& -5L^{\frac{1}{4}} \lceil z_1-v_0 \rfloor^{\frac{3}{4}} \teq v_1 \cr
\dz_2 &=& -3L^{\frac{1}{3}} \lceil z_2-v_1 \rfloor^{\frac{2}{3}} \teq v_2 \cr
\dz_3 &=& -1.5L^{\frac{1}{2}} \lceil z_3-v_2 \rfloor^{\frac{1}{2}} \teq v_3 \cr
\dz_4 &=& -1.1L \sgn(z_4-v_3)
\end{eqnarray}
where $\lceil z \rfloor^p = \abs{x}^p \sgn(x)$ and $L$ is a positive design constant. The estimate of $a^{(i)}$ is $z_i$. It is hard to satisfy the condition that the settling time of $z_4$ is $0.1$s without using fairly large $L$ value. Thus, we has chosen $L=3\times 10^7$, which leads to severe chattering in $z_4$ as shown in fig. \ref{sim31}.
%
%
%%%%%%%%%\%%%%%%%%%\%%%%%%%%%\%%%%%%%%%\%%%%%%%%%\%%%%%%%%%\%%%%%%%%%\%%%%%%%%%
%
%
%
%
%
\section{Conclusion}
%
%
%
%
%
%%%%%%%%%\%%%%%%%%%\%%%%%%%%%\%%%%%%%%%\%%%%%%%%%\%%%%%%%%%\%%%%%%%%%\%%%%%%%%%
A novel SD that has considerably simple dynamics is proposed. Under the assumption that time-derivatives of the signal are norm-bounded, it is shown that estimation error is convergent to the zeros asymptotically. The estimated derivative shows neithor chattering nor peaking pheonomenon and tracks the desired value exactly after finite transient period. A 1st-order diffentiator is firstly proposed and, by connecting this differentiator in series, higher-order derivatives are also available. Simulation results depict that the proposed differentiators show extreme performaces compared to the widly used previous differntiators such as HGO or HOSM differentiator.
%\section*{Acknowledgment}


\ifCLASSOPTIONcaptionsoff
  \newpage
\fi



%\bibliographystyle{IEEEtran}
\bibliographystyle{IEEEtran}
\bibliography{../bib/mypaper,../bib/paper,../bib/book}

%\begin{thebibliography}{1}
%\end{thebibliography}


%\end{IEEEbiography}


%\newpage



%\vfill


\end{document}